\documentclass[pra,
twocolumn,
superscriptaddress,showpacs,floatfix,nofootinbib]{revtex4}
\usepackage{epsfig,amsmath,amsfonts,amssymb}
\usepackage{graphicx}
\usepackage[utf8]{inputenc}
\usepackage{boxedminipage}
\usepackage{epsfig,psfrag}
\usepackage{amsmath}
\usepackage{mathbbol, amsfonts}
\usepackage{lscape}
\usepackage{fancybox}
\usepackage{amsfonts,amssymb}   
\usepackage{color}
\usepackage{subfig}
\newcommand{\beq}{\begin{equation}}
\newcommand{\eeq}{\end{equation}}

\newcommand{\beqa}{\begin{eqnarray}}
\newcommand{\eeqa}{\end{eqnarray}}
\newcommand{\ba}{\begin{align}}
\newcommand{\ea}{\end{align}}

\usepackage{wrapfig}

\begin{document}
\title{Time-Optimal Transport of a Harmonic Oscillator: Analytic Solution }

\author{ Gerhard C. Hegerfeldt}
\affiliation{Institut f\"ur Theoretische Physik, Universit\"at G\"ottingen,
Friedrich-Hund-Platz 1, D-37077 G\"ottingen, Germany}

\begin{abstract}
Motivated by the experimental transport of a trap with a quantum mechanical system modeled as a harmonic oscillator (h.o.)  the corresponding classical problem  is investigated.  
Protocols for the fastest possible transport of a classical h.o. in a wagon over a distance $d$ are derived where both initially and finally the  wagon is at rest  and the h.o. is  in its equilibrium position and also at rest. The acceleration of the wagon  is assumed to be bounded. 
For fixed oscillator frequency $\Omega$ it is shown that  there are in general  three switches in the acceleration and  for special values of  $\Omega$ only one switch. In the latter case the optimal transport time is $T_{\rm abs}$, that of a wagon without oscillator. 
The optimal transport time and the switch times are determined. It is shown that in some cases it is advantageous to go {\em backwards} for a while.
In addition a time-dependent  $\Omega(t)$, bounded by $\Omega_\pm$, is allowed.
In this case the behavior depends sensitively on $\Omega_\pm$ and is spelled out  in detail. In particular, depending on $\Omega_\pm$, $T_{\rm abs}$ may be obtained in continuously many ways.

\end{abstract}
%\pacs{45.10.Db, 02.30Yy, 03.65.Ca}
\maketitle

\section{Introduction}\label{section1}

Adiabatic processes may serve to  transform an initial state of a system to a proscribed final state. Such processes, however, are very slow and, in principle, infinitely slow. Protocols for speeding up the time development have been introduced in the past, with numerous applications in quantum optics \cite{0,1,01,02,2,3,4,5,6,7,8,9,10,10a,11,12,12a,13,14,14a,15,16} and to classical systems, e.g. cranes \cite{cranes}. These methods include `shortcuts to adiabadicity' (STA) \cite{0,1,01,02,2,3,4,5,6,7},  `counterdiabatic' approaches \cite{8,9,10} and the `fast-forward' approach \cite{11,12,12a,13}.  In general the above mentioned protocols yield a speed-up, but not necessarily the fastest possible time development.
Other methods are combinations with control theory \cite{pont,hock,Boscain}, cf. e.g. \cite{kosloff2017,14,stefmuga2011}. While a time development as fast as possible is often desired, other considerations like robustness and further conditions may prolong the resulting time duration.

A particular example is the efficient transport of ultra cold atoms and ions by moving the confining trap. An atom or ion in a harmonic trap can be treated to good approximation as a quantum harmonic oscillator. For harmonic traps efficient protocols have been investigated with STA and the invariant-based inverse engineering method to obtain  transitionless evolutions under imposed constraints, faster than by an adiabatic process \cite{3,stefmuga2011}.
It is therefore natural to ask how fast the transport  of a quantum harmonic oscillator can be made. This depends of course on the particular question one is interested in, for example a time-optimal transport a a harmonic oscillator under additional conditions.

Insight for the quantum case may be obtained by asking  the same question  for a classical harmonic oscillator. Therefore in this paper the time-optimal transport of  a classical harmonic oscillator will be investigated.

\begin{figure}[h]
\begin{center}
%\scalebox{0.26}[0.26]{\includegraphics{fig2.eps}}
\scalebox{0.45}[0.45]{\includegraphics{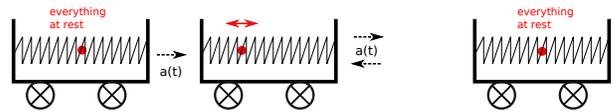}}
\caption{\label{wagon1} Oscillating mass $m$ attached to a spring in an accelerated wagon}
\end{center}
\end{figure}

Consider a classical one-dimensional harmonic oscillator (h.o.) without friction in the center of a long wagon, such as depicted in Fig. \ref{wagon1} where a small mass $m$ is attached to a spring on the wagon.  When the wagon is accelerated the h.o. will start to perform oscillations. In this case the frequency $\Omega$ of the h.o. depends on the spring constant and on $m$.

The problem to be investigated is the following:

(i) Initially  the wagon is at rest  and the h.o. is  in its equilibrium position, also at rest.

(ii) Then the wagon undergoes an acceleration $a(t)$, where  $a(t)$ can vary between  $\pm a_{\rm max}$, until it has traveled a prescribed distance d.

(iii) Upon arrival at the end point the system should again be in its initial state, i.e. the wagon should be at rest, and  the h.o. should again be in its equilibrium position and  at rest.

The questions to be answered here are: Is this achievable, and if so what is the shortest time possible? Can this time be further lowered by allowing the h.o. frequency $\Omega$ to be time dependent, i.e. $\Omega(t)$? Both questions will be answered in the affirmative.

The plan of the paper is as follows. First, in Section \ref{section2}, a fixed oscillator frequency  will be considered, examples will be given  and a complete solution of the problem and an explicit protocol for fixed $\Omega$ will be formulated. In Section \ref{section3} detailed proofs are provided. In Section \ref{section4} the case of a time-dependent oscillator frequency  is treated where $\Omega(t)$ satisfies $\Omega_- \leq \Omega(t) \leq \Omega_+$, with arbitrary $\Omega_\pm$. The results and protocols will be seen to depend critically on the particular choice of $\Omega_\pm$. Finally, in Section \ref{discussion}  the results are summarized and discussed.

\section{  Optimal protocol for fixed oscillator frequency}\label{section2}
We consider a classical one-dimensional harmonic oscillator  on a long wagon. The position of the h.o. (i.e. mass point) relative to the wagon center will be denoted by $x_{\rm h}$ and the position of the wagon center in the external rest frame by  $x_{\rm w}$. When the wagon is accelerated with acceleration $a(t)$, the mass point additionally experiences the corresponding inertial force $-m a$ in the rest frame of the wagon so that one has
\beqa \label{2.1}
\ddot{x}_{\rm h} & =& - \Omega^2 x_{\rm h}-a~\\
\ddot{x}_{\rm w} &=& a~.\nonumber
\eeqa
It is assumed that $a(t)$ can vary between  $\pm a_{\rm max}$.

\begin{figure}[h]
\begin{center}
%\scalebox{0.26}[0.26]{\includegraphics{fig2.eps}}
\scalebox{0.9}[0.9]{\includegraphics{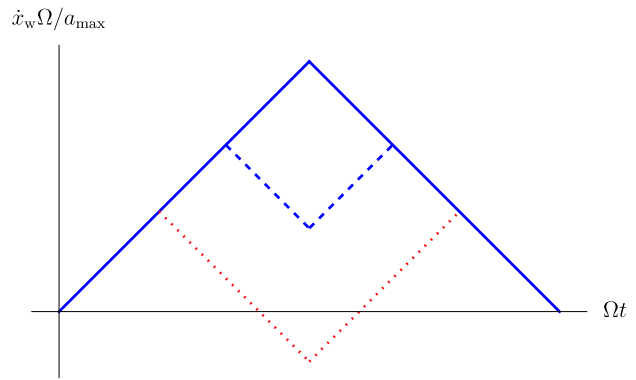}}
\caption{\label{wagon2} Typical wagon velocities for the acceleration alternating between$\pm 1$. Solid curve: No oscillator present and  Example 2 with resonant $\Omega$. Dashed and dotted curves: General $\Omega$.  For the dotted curve the wagon velocity becomes partially negative, i.e. the wagon moves {\em backwards} for some time.}
\end{center}
\end{figure}

{\em Example} 1. With no h.o. present, to move a wagon a distance $d$ in shortest time, with initial and final velocity equal to zero, it is optimal to accelerate with $a_{\rm max}$  for half the distance and then decelerate with $-a_{\rm max}$ \cite{hock}  (cf. solid line in Fig.\ref{wagon2}). The corresponding time $T_{\rm abs}(d)$, $T_{\rm abs}^2 = 4\,d/a_{\rm max}$, can at most be achieved, but not undercut, if a h.o. in the wagon is to be initially and finally at rest in its equilibrium position.

{\em Example} 2. For special 'resonant values' of $\Omega$ this time can indeed be achieved, e.g. for  
\beqa \label{resonance}
\begin{split}
  \Omega &=n\, \Omega_{\rm res}(d)~,~~~~~~n=1,2,\cdots ~~~~\\
  \Omega_{\rm res}(d)&= \sqrt{4\pi^2 a_{\rm max}/d}= 4\pi/T_{\rm abs}~. 
\end{split}
\eeqa
To see this consider $n=1$. Initially, the wagon and h.o. are at rest. Upon accelerating the wagon by $a_{\rm max}$ the h.o. experiences, in the wagon frame, the additional inertial force $-ma$ and starts to move to the left. During the time $T_{\rm abs}/2$ it has just performed a single oscillation, has returned to its initial position in the wagon and is at rest relative to the wagon. In this instant, the acceleration of the wagon is reversed, the h.o. starts moving to the right and at a further time duration of $T_{\rm abs}/2$ is back at rest at the initial position, with the wagon at rest and having traveled the distance $d$. For $n>1$ one has correspondingly more oscillations.

For fixed $\Omega$ a protocol to obtain the unique optimal transport time is constructed as follows.\\[.2cm]
{\em 
(i) For given $d$  determine the unique optimal time $t_{\rm f}$  by the equation
\beq \label{2.3}
d = \frac{1}{4}a_{\rm max}\, t_{\rm f}^2 \,[1 - \frac{8}{(\Omega\,t_{\rm f}) ^2}\left(\arccos(\cos^2(\Omega\, t_{\rm f}/4))\right)^2]~.
\eeq
(ii) With wagon and oscillator at rest at $t = 0$, accelerate with $a_{\rm max}$  until time  $\frac{1}{2}t_{\rm f} - t_1$  where $t_1$, $0\leq \Omega t_1 \leq\pi/2 $,  is given by
\beq\label{2.4}
t_1 =\frac{1}{\Omega}\arccos(\cos^2(\Omega\, t_{\rm f}/4))~.
\eeq
(iii) Decelerate with $-a_{\rm max}$ until time $\frac{1}{2}t_{\rm f}$.
\\
(iv) Accelerate with $a_{\rm max}$ until time  $\frac{1}{2}t_{\rm f} + t_1$.
\\
(v) Finally decelerate with  $-a_{\rm max}$  until time $t_{\rm f}$. }
\\

Typical wagon velocities are depicted in Fig.~\ref{wagon2}.
At the end the wagon is obviously at rest. The oscillator may perform several oscillations. That finally it is also again  at rest and in its equilibrium position will be shown at  the end of this section.  In the next section it will be shown that $t_{\rm f}$ is indeed the unique optimal time. The above protocol has a certain symmetry; there may, or may not, be  other, nonsymmetric, protocols which lead to the same unique optimal time.

Note that $t_1 = 0$ if $\Omega\, t_{\rm f} = 4 n \pi$,  $ n=1, 2, \cdots$, which recovers Example 2 with $t_{\rm f}= T_{\rm abs} $. 
If $t_1>\frac{1}{4}t_{\rm f}$ the wagon velocity temporarily becomes negative (dotted curve in Fig. \ref{wagon2}), i.e. then it is advantageous to go {\em backwards} for a while. From Eqs. (\ref{2.3}, \ref{2.4}) this is seen to happen if
\beq\label{2.5}
\Omega^2 < \frac{1}{4}\, \Omega_{\rm res}(d)^2 ,
\eeq
i.e. for small oscillator frequency. However, it can easily be shown that the backward motion will not go back as far as the original starting position of the wagon.

\begin{figure}[h]
\begin{center}
%\scalebox{0.26}[0.26]{\includegraphics{fig2.eps}}
\scalebox{0.9}[0.9]{\includegraphics{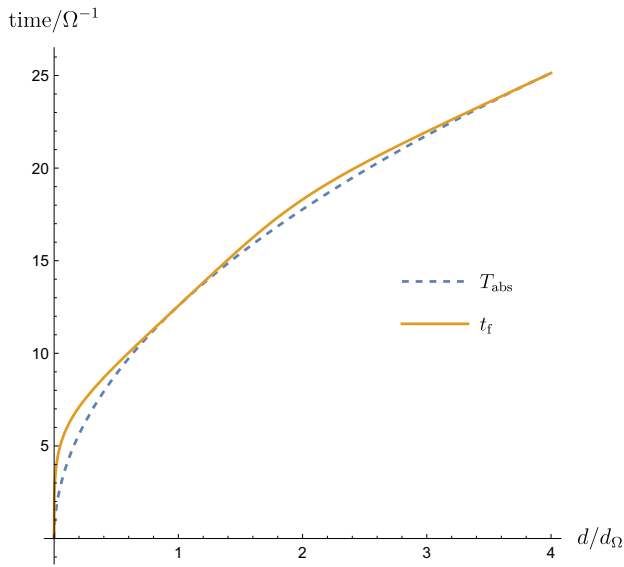}}
\caption{\label{time-d} Solid curve: Optimal transport time $t_{\rm f}$  as a function of distance $d$ in units of $d_\Omega= 4\pi^2a_{\rm max}\,\Omega^{-2}$, for fixed $\Omega$. Dashed curve: $T_{\rm abs}(d)$  (without oscillator). For $d/d_\Omega =1,~2^2,~\cdots$ the times coincide.}
\end{center}
\end{figure}

If one plots $d$ as a function of $t_{\rm f}$ in Eq.(\ref{2.3}) then $t_{\rm f}$ as a function of $d$ is given by reflecting it at the diagonal.
In dimensionless scaled variables, the solid curve in Fig. \ref{time-d} displays   $\Omega\, t_{\rm f}$ as a function of $d/d_\Omega$ where $d_\Omega = 4\pi^2a_{\rm max}\,\Omega^{-2} $ is the distance for which $\Omega$ is resonant, i.e. $\Omega_{\rm res}(d_{\Omega})= \Omega $.
The dashed curve is the corresponding  $ T_{ \rm abs}(d)$. Note that at $d/d_\Omega =n^2, n= 1,2,\cdots$ the two transport times coincide, which is again Example 2. 

\begin{figure}[h]
\begin{center}
%\scalebox{0.26}[0.26]{\includegraphics{time-om.eps}}
\scalebox{0.9}[0.9]{\includegraphics{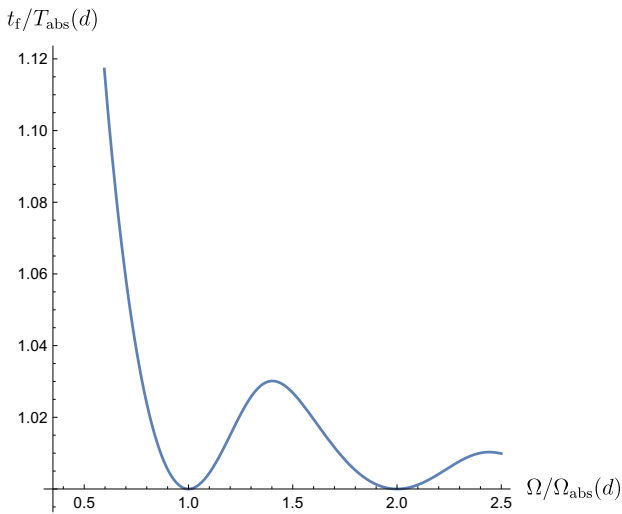}}
\caption{\label{time-om} Fixed $d$: Optimal transport time $t_{\rm f}$  in units of $T_{\rm abs}(d)$ as a function of  $\Omega$ in units of $\Omega_{\rm abs}(d)$.  }
\end{center}
\end{figure}

For fixed $d$, one can also obtain  $t_{ \rm f}$ as a function of $\Omega$ from Eq. (\ref{2.3}). In dimensionless scaled variables the result is plotted in  
Fig. \ref{time-om}. It is seen that $t_{ \rm f}$ diverges for $\Omega \to 0$. This can be made more explicit by expanding Eq. (\ref{2.3}) in terms of $\Omega\, t_{\rm f}$. A short calculation gives, in  dimensionless scaled variables,
\beq \label{2.6}
t_{ \rm f}/ T_{ \rm abs}(d)\approx \{6/\pi^2\}^{1/4}\,(\Omega/\Omega_{\rm abs}(d))^{-1/2}.
\eeq
Replacing 6 by 5.3 in Eq.(\ref{2.6}) one obtains an excellent approximation for $t_{ \rm f}/ T_{ \rm abs}(d)$ in the range $0.05\leq \Omega/\Omega_{\rm abs}(d) \leq 0.7$.

{\em Protocol evaluation.} For the oscillator time-development   Eq. (\ref{2.1}) has to be evaluated with $a = \pm a_{\rm max}$. This is conveniently done in the complex plane. With
\beq
z = x_{\rm h} + {\rm i}\, \Omega^{-1}{\dot x}_{\rm h} \pm a_{\rm max}/\Omega^2
\eeq
one finds $\dot z = -{\rm i} \Omega \,z$ and thus $z(t) = \exp[-{\rm i}\Omega (t-t_0)\,z(t_0)$. Hence
\ba \label{complex}
x_{\rm h}(t) &+ {\rm i}\, \Omega^{-1}{\dot x}_{\rm h}(t)=  \exp[-{\rm i}\Omega (t-t_0)]\\
&\cdot( x_{\rm h}(t_0) + {\rm i}\, \Omega^{-1}{\dot x}_{\rm h}(t_0)\pm a_{\rm max}/\Omega^2) \mp  a_{\rm max}/\Omega^2. \nonumber
\end{align}
In the complex plane the right-hand side corresponds to a clock-wise rotation of $ x_{\rm h}(t_0) + {\rm i}\, \Omega^{-1}{\dot x}_{\rm h}(t_0)$ by the angle $\Omega (t-t_0)$ around the point $-a_{\rm max}/\Omega^2$ and $a_{\rm max}/\Omega^2$, respectively.

In the protocol one starts with $x_{\rm h}(0)=0$ and $\dot x_{\rm h}(0)=0$ and rotates clock-wise around $-a_{\rm max}/\Omega^2$, then around  $a_{\rm max}/\Omega^2$, then again around  $-a_{\rm max}/\Omega^2$ and finally around  $a_{\rm max}/\Omega^2$.
\begin{figure}[h]
\begin{center}
%\scalebox{0.26}[0.26]{\includegraphics{fig2.eps}}
\scalebox{0.9}[0.9]{\includegraphics{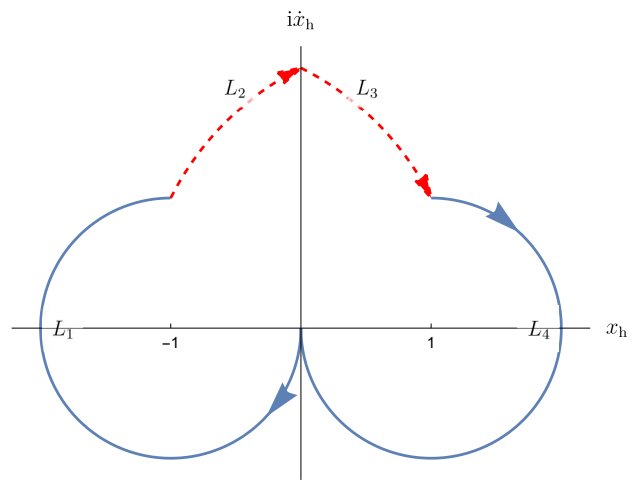}}
\caption{\label{complexrot} Time-development of $x_{\rm h}$ in complex phase-space  for $\Omega =1$, $a_{\rm max}=1$, $d=2.82\, \pi^2$, $t_{\rm f}=3.41\, \pi$, and $t_1=.205\,\pi$. Starting at the origin, i.e. equilibrium position and at rest, there is first a rotation around -1, then around 1, then around -1 and finally again around 1, back to the origin.}
\end{center}
\end{figure}
Analytically this gives for the first two rotations
\ba\label{2.11}
\zeta_1&\equiv x_{\rm h}(t_{\rm f}/2 - t_1) + {\rm i}\, \Omega^{-1}{\dot x}_{\rm h}(t_{\rm f}/2 - t_1)\nonumber\\
&= \exp[-{\rm i}\Omega (t_{\rm f}/2 - t_1)]\,a_{\rm max}/\Omega^2-a_{\rm max}/\Omega^2\nonumber\\
\zeta_2 &\equiv  x_{\rm h}(t_{\rm f}/2) + {\rm i}\, \Omega^{-1}{\dot x}_{\rm h}(t_{\rm f}/2 )\nonumber\\
&=\exp[-{\rm i}\Omega t_1](\zeta_1-a_{\rm max}/\Omega^2)+a_{\rm max}/\Omega^2.
\end{align}
$x_{\rm h}(t_{\rm f}/2)$ is the real part of $\zeta_2$ and  one finds 
\ba\label{2.12}
x_{\rm h}(t_{\rm f}/2)&=2\, a\,_{\rm max}/\Omega^2\,(\cos^2(\Omega\,t_{\rm f}/4)) - \cos (\Omega\,t_1))\nonumber\\
&= 0
\end{align}
by Eq. (\ref{2.4}),  i.e. $\zeta_2$ lies on the imaginary axis. The corresponding trajectories in the complex plane correspond to the two curves in the left half-plane in Fig. 5. By the symmetry of the protocol the next two steps give the two curves in the right half-plane where the last one ends again at the origin. This follows of course also analytically. Hence after the final step the oscillator is again at rest in its equilibrium position. Thus the protocol satisfies the initial and final conditions.

\section{Proof of Optimality for fixed $\Omega$}\label{section3}
First the equivalent {\em converse} problem will be considered: Finding  the longest distance $d$ for a given time duration $t_{\rm f}$  under the conditions (i) - (iii) in Section \ref{section1} and a corresponding protocol.

{\em Symmetry.} Consider some given $t_{\rm f}$ and $d$. In the following it is convenient to let time run from $-\frac{1}{2} t_{\rm f}$ to $\frac{1}{2}t_{\rm f}$. Let $x_{\rm h}$ and $x_{\rm w}$ satisfy Eqs. (\ref{2.1}) for some $a(t)$ and the boundary conditions at $\pm \frac{1}{2} t_{\rm f}$. Then $\frac{1}{2}(x_{\rm h}(t)-x_{\rm h}(-t))$ and $\frac{1}{2}(x_{\rm w}(t)-x_{\rm w}(-t)+d)$ satisfy Eqs. (\ref{2.1}) with $a(t)$ replaced by $\frac{1}{2}(a(t)-a(-t))$ and the same boundary conditions. Hence without loss of generality one can assume that  $x_{\rm h}$ and  $a$ are {\em anti-symmetric} while  $\dot x_{\rm h}$ and  $\dot x_{\rm w}$ are {\em symmetric} under time reversal.

{\em Scaled variables.} We go over to dimensionless scaled variables. We choose some fixed length unit $d_0$ and put 
\beqa \label{3.1}
\Omega_0^2&= a_{\rm max}/d_0\hspace{1.5cm} \omega&= \Omega/\Omega_0\nonumber\\
\tau&=\Omega_0 t \hspace{1.5cm} u(\tau) &= a(t)/a_{\rm max} \nonumber \\
{\xi}_1(\tau)&=  x_{\rm h}(t)/d_0 \hspace{1cm}\xi_2(\tau) &=\frac{d}{d\tau}\xi_1(\tau)\nonumber \\
\xi_3(\tau)&= x_{\rm w}(t)/d_0  \hspace{1.3cm} \xi_4(\tau)&=\frac{d}{d\tau}\xi_3(\tau)
\eeqa
so that $u(\tau)$ can vary between $\,-1$ and 1. Then one obtains
\beqa \label{3.2}
\ddot\xi_1\equiv \frac{d^2}{d\tau^2}\xi_1&=& - \omega^2 \xi_1 - u(\tau)\\
\ddot\xi_3 &=&u(\tau) ~.\nonumber
\eeqa
For fixed $\Omega$ and a suitable $d_0$ one can assume $\Omega_0 =\Omega$ and then $\omega=1$.

{\em Pontryagin Maximum (or Minimum) Principle} (PMP) \cite{pont,hock,Boscain}.
This is a far-reaching generalization of the calculus of variations and regarded as a milestone in control theory. A simple example is  a car moving in shortest time from standstill at A to standstill at B, under the only condition that the time-dependent acceleration resp. deceleration (the 'control') is bounded, but not  necessarily continuous.

The PMP serves to determine necessary conditions for an optimal control function $ u^\ast (t)$ (or possibly several control functions) which minimizes
a given cost function $J$ of the form $J= \int^T_0 L(u(\tau),...) d\tau$, where $L$ is a function of the control $u(\tau)$ and some state functions $\xi_i$ and their derivatives.
For the present distance-optimal control problem, one can take $L=\xi_4$ since $J = \int^T_0 \dot\xi_3 d\tau$ is the (scaled) distance. To minimize it, the PMP considers a control
Hamiltonian $H_ c$,
\ba
\label{3.3}
H_c = -L + &  p_1 \dot\xi_1 + p_2 \dot\xi_2 + p_3 \dot\xi_3+p_4\dot\xi_4,
\end{align}
where one inserts $\dot\xi_i$ from Eqs. (\ref{3.1}-\ref{3.2}) and where the adjoint states $p_i$ are Lagrange multipliers which can not all be identically zero. Then, for an extremal control $u(\tau) = u*(t)$,    
Hamilton's equations
\ba 
\dot p_i = - \partial H_c/\partial \xi_i,
\hspace{1.5cm}
\dot \xi_i= \partial H_c /\partial p_i 
\end{align}
hold.
For almost all $-\tau_{\rm f}/2 \leq \tau \leq \tau_{\rm f}/2$, the function $H_c (p_i(t ), \xi_i(t), u(t))$ attains
its maximum at $u(t) = u^\ast (t)$, and $H_c ={\rm const}$.
For simplicity we omit the asterisk on $u^\ast$. Inserting for $\dot\xi_i$,  $H_ c$ becomes
\ba
\label{3.4}
H_c = -\xi_4 +  p_1\xi_2 + p_2 ( -\omega^2 \xi_1 - u) + p_3 \xi_4 + p_4 u~.
\end{align}\\

From the term $(p_4-p_2)\,u$ it follows that for a maximum one has to choose $u(\tau)= 1$ if  $p_4-p_2 > 0$ and -1 if  $p_4-p_2 < 0$. When  $p_4-p_2 = 0$, or more precisely, when $p_4-p_2$ changes sign, there is a switch from $\pm 1$ to $\mp 1$ in $u$. 
Hamilton's equations become
\ba \label{3.5}
\dot p_1 &= \omega^2 p_2, \hspace{2cm}     \dot p_2 = -p_1 \nonumber\\
\dot p_3 &=0, \hspace{2.55cm}      \dot p_4 = -p_3+1
\end{align}
The solutions are
\ba \label{3.6}
p_2(\tau) &=A \cos\tau + B\sin\omega\tau,\hspace{1cm} p_1 =-\dot p_2 \nonumber\\
p_3 &=c_3,\hspace{2cm}  p_4 =( -c_3+1)\,\tau + c_4 
\end{align}
where $A$, $B$, $c_3$, and $c_4$ are constants. If $p_4-p_2\equiv 0$ in some extended interval, then  $p_4 = p_2\equiv 0$, by linear independence. Therefore it is not possible to have $u\equiv 0$ and $\xi_4 \equiv const$ in some extended interval so that there are only isolated switches. Hence, by anti-symmetry of $u$, there is a switch at $\tau =0$, i.e. $-p_2(0)+p_4(0)=0$, and thus $A=c_4$. By the boundary conditions on $\xi_{\rm i}$ at $\pm\tau_{\rm f}/2$ only the terms containing $u$ remain in $H_c$ which by antisemitic of $u$ lead to two equations and to
\beq \label{3.6a}
A\,(\cos(\omega\tau_{\rm f}/2)-1)=0~.
\eeq
Thus either $A=0$ or $\omega\tau_{\rm f}=4\pi n$. In the latter case the situation is analogous to Example 2,  i.e. the h.o. can perform $2n$ complete oscillations and the optimal distance is the same as without oscillator.  We can therefore assume $A=c_4=0$. For $\omega\tau_{\rm f}\neq 4 \pi n$ there are at least two switches of $u$ and therefore  $ B\neq 0$ since otherwise $-c_3 + 1=0$, $c_3=1$, and $\xi_4\equiv$ const. The explicit values of $B$ and $c_3$ are not needed, they can in principle be calculated at the end; it suffices to discuss the cases $B<0$ and $B>0$.

{\em Note:}  From the remark after Eq. (\ref{3.4}) it follows that $u(\tau) = 1$ when the line $p_4 (\tau)$ lies {\em above} the sine curve $p_2(\tau)$ and $u(\tau)=-1$ when it lies {\em below}.

{\em Case} $B<0$. (i) Single switch for $\tau <0$, at $-\tau_1$, say. Then the line $p_4(\tau)$, denoted by $L_1$ in Fig.~6, intersects with the -sine curve $p_2(\tau)$ once.

\begin{figure}[h]
\begin{center}
%\scalebox{0.26}[0.26]{\includegraphics{time-om.eps}}
\scalebox{0.9}[0.9]{\includegraphics{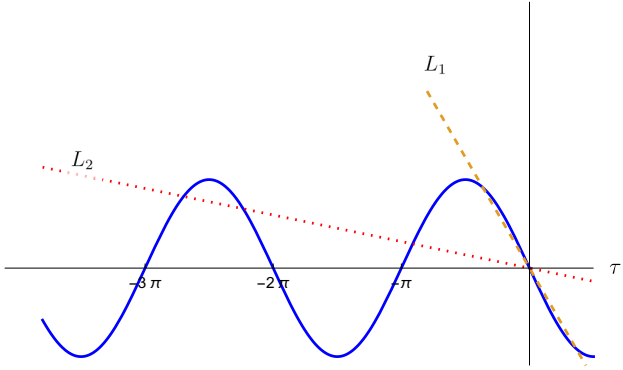}}
\caption{\label{B1}Case $B<0$. With $\omega =1$. $L_1$ and $L_2$ denote possible lines for $p_4(\tau)$. Their intersections with $p_2(\tau)$ (-sine curve) are possible switching points. In regions where $p_4(\tau)$ is above $p_2(\tau)$ one has acceleration, otherwise deceleration. Only $L_1$ with a single switch is optimal. }
\end{center}
\end{figure}

The analog of Eqs. (\ref{2.11}) for $\xi+{\rm i}\omega^{-1}\dot \xi$  in the scaled variables, now with initial time  -$\tau_{\rm f}/2$  and final time 0 yields
\ba\label{3.7}
\xi_1(0) = \cos(\omega\tau_{\rm f}/2) - 2 \cos(\omega\tau_1) +1.
\end{align}\label{3.8}
From the anti-symmetry of $\xi_1$ one has $\xi_1(0)=0$, and from this one obtains
\beq\label{tau1}
 \cos \omega\tau_1=\cos^2(\omega\tau_{\rm f}/4)
\eeq
with $-\pi/2\omega<-\tau_1<0$. Thus line $L_1$ in Fig.~\ref{B1} is typical in this case, while line $L_2$ is not possible.

(ii) If there are two or more switches for $\tau <0$, e.g. if $p_4(\tau)$ is given by line $L_2$ in Fig.~\ref{B1}, then the last deceleration period before $\tau = 0$ is longer than $\pi/2\omega$. Hence the total acceleration time is less than in (i) and the distance traveled by the wagon during $\tau_{\rm f}$ is less than that in (i).
Hence for $B<0$ there is only a single switch for $\tau < 0$. 

\begin{figure}[b]
\begin{center}
%\scalebox{0.26}[0.26]{\includegraphics{time-om.eps}}
  \scalebox{0.9}[0.9]{\includegraphics{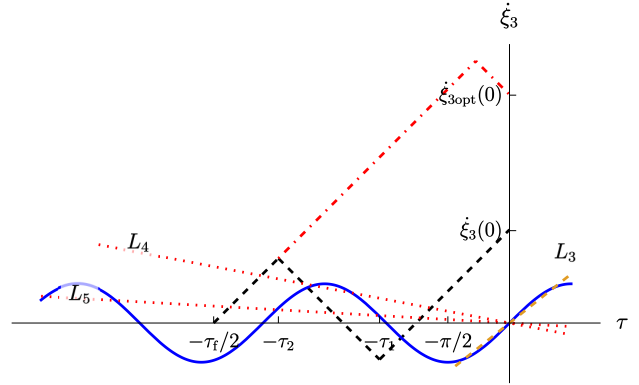}}
  \caption{\label{B2}Case $B>0$. With $\omega = 1$.
    $L_3$, $L_4$ and $L_5$ denote possible lines for $p_4(\tau)$. Their intersections with $p_2(\tau)$ (sine curve) are possible switching points. Dashed: $\dot\xi_3$ with 2 intersection points $-\tau_1$ and $-\tau_2$. Dotdashed: $\dot\xi_{3 \rm opt}$ from case $B<0$. For $\tau_2 -\tau_1 > \pi/2$ one has $\dot\xi_{\rm opt}>  \dot\xi$.
    $L_3$ is typical for the optimal backwards motion.}
\end{center}
\end{figure}

{\em Case} $B>0$. From Fig. \ref{B2} this is case $B<0$ reflected at the $\tau$ axis, with $u=\pm 1$ interchanged and thus positive wagon distances for $B<0$ now  become negative. But there might also be negative distances for $B<0$, corresponding to positive distances for $B>0$, and therefore a more detailed discussion is required. Here we use $\omega=1$.

(i) Single switch for $\tau < 0$: As for $B<0$ there is only a single solution for fixed $\tau_{\rm f}$, and this is the corresponding optimal backward motion,  with $p_4(\tau)$ typically given by $L_3$ in Fig.~\ref{B2}.

(ii) Exactly two switches for $\tau <0$. Typical for this would be lines $L_4$ and $L_5$  in Fig. \ref{B2}, with  switches at
$-\tau_2< -\tau_1<0$, say. \\
a) Case $\tau_2-\tau_1 > \pi/2$.\\
From Fig.~7  one easily finds $\dot\xi_3(0) = \tau_{\rm f}/2 - 2 (\tau_2 -\tau_1) < \tau_{\rm f}/2  -\pi$ while, from case $B<0$, $\dot\xi_{3\rm opt}\geq \tau_{\rm f}/2  -\pi$ since here the switching point lies to the right of $-\pi/2$. Hence in case $B<0$ the distance  is larger.\\
b) Case $\tau_2-\tau_1< \pi/2$.\\ 
This will be shown to be incompatible with the boundary conditions on the h.o..
  One has $\xi_1(0)=0$, by anti-symmetry, while $\dot\xi_1(0) \equiv \lambda $ is unknown.
 Reversing the time development from $\tau=0$ to $\tau=-\tau_2$ one obtains
 \ba
  \xi_1(-\tau_1)+{\rm i}\dot \xi_1(-\tau_1)&=\exp[-{\rm i} \tau_1]\{{\rm i}\lambda +1\}-1\nonumber\\
  \xi_1(-\tau_2)+{\rm i}^{-1}\dot \xi_1(-\tau_2)&  = \nonumber\\
 &\hspace{-3.5cm} \exp[{\rm i} (-\tau_1+\tau_2)]\{\xi_1(-\tau_1)+ {\rm i}\dot\xi_1(-\tau_1)-1\} +1
  \nonumber\\
  &\hspace{-3.5cm}
  =\exp[{\rm i} (-\tau_1+\tau_2)\{\exp[{\rm i}\tau_1]({\rm i}\lambda +1)-2\}
  +1
 \end{align}
 Since this must lie on the circle around $-1$ passing through 0, upon adding $1$ the rhs  becomes a number of modulus  $1$:
\ba 
1& =|\exp[{\rm i} (-\tau_1+\tau_2)]\nonumber
\{\exp[{\rm i}\tau_1]({\rm i}\lambda + 1)-2\} +2|\nonumber\\
&= |{\rm i} \lambda +1 -2\exp[-{\rm i}\tau_1] + 2\exp[-{\rm i}\tau_2]|
\end{align}
Hence the modulus of the real part,
\ba\label{3.Re}
|1-2\cos\tau_1 + 2\cos\tau_2|,
\end{align}
must be less than, or equal to, $1$. However, from Fig. \ref{B2},  one has  $- 3\pi/2 <-\tau_1<-\pi$ and so $\cos\tau_1 < 0$. For $-2\pi <-\tau_2< -3\pi/2 $ one has $\cos\tau_2 > 0$ while for $ -3\pi/2<-\tau_2<- \pi$ one has $-2\cos\tau_1 + 2\cos\tau_2>0$. Hence the bracket in Eq. (\ref{3.Re}) is larger than 1,  a contradiction. Thus this case can not occur.\\

(iii) Three or more switches for $\tau < 0$: A typical line is $L_5$ in Fig.~\ref{B2}. From Fig.\ref{B2} it is evident that the area under the curve (i.e. distance) decreases.

As a consequence, case $B>0$ is not possible and case $B<0$ (i) gives the unique optimal distance for given $\tau_{\rm f}$ and fixed $\omega$ in scaled variables. This distance is easily calculated to be $\tau_{\rm f}^2/4 - 2 \tau_1^2$, with $\tau_1$, $0\leq \tau_1\leq \pi/2$, given by Eq.~(\ref{tau1}).
In the original variables one has
\beq\label{d}
d= \frac{1}{4}a_{\rm max}t_{\rm f}^2 - 2a_{\rm max}t_1^2.
\eeq
Going back to the original problem one obtains the protocol of Section \ref{section2}.\\

\section{Protocols for time-dependent oscillator frequency }\label{section4}

In this case one allows in addition to $a(t)$ also $\Omega (t)$ to be time-dependent and seeks a minimal transport time $t_{\rm f}$ for a distance $d$ under the condition that the wagon is initially and finally at rest and the oscillator is at rest in its equilibrium position.
This situation is more complicated. If there are no bounds on $\Omega$ then for $\Omega \to \infty$ one obtains the absolute minimal time as without oscillator. Therefore, in addition to $|a(t)|\leq a_{\rm max}$ one imposes bounds
\beq
0 \leq \Omega_- \leq \Omega(t) \leq \Omega_+ <\infty .
\eeq
If a 'resonant value'  from Eq. (\ref{resonance}) lies in this interval then, from Example 2, one chooses this value for  $\Omega$ and then obtains the absolute minimal time.

{\em Distance optimization}.  Again we first consider the equivalent problem of finding a protocol that maximizes the distance $d$ for given time $t_{\rm f}$ and let time run from $-\frac{1}{2} t_{\rm f}$ to $\frac{1}{2}t_{\rm f}$. We will seek solutions that satisfy the same symmetry properties as in Section \ref{section3}, i.e. we assume that  $\Omega(t)$ is {\em symmetric}.

The same scaled variables as in Eq. (\ref{3.1}) are used. Introducing 
\beq\label{4.1}
u_1(\tau)\equiv \omega^2(\tau)
\eeq
as a second control variable, Eq. (\ref{3.2}) reads
\beqa \label{4.2}
\ddot\xi_1\equiv \frac{d^2}{d\tau^2}\xi_1&=& - u_1(\tau) \xi_1 - u(\tau)\\
\ddot\xi_3 &=&u(\tau) ~.\nonumber
\eeqa

The condition on $\Omega(t)$ becomes  $\omega_-^2\leq u_1(\tau) \leq \omega_+^2$. The control Hamiltonian for the PMP now reads
\ba
\label{4.3}
H_c = -\xi_4 +  p_1\xi_2 + p_2 ( -u_1 \xi_1 - u) + p_3 \xi_4 + p_4 u~.
\end{align}\\
As before it follows that for a maximum one has to choose $u(\tau)= 1$ if  $p_4>p_2 $ and -1 if  $p_4<p_2 $. When  $p_4-p_2 = 0$, or more precisely, when $p_4-p_2$ changes sign, there is a switch from $\pm 1$ to $\mp 1$ in $u$.  Similarly, $u_1=\omega_+^2$ if $p_2\xi_1 < 0$, and $u_1=\omega_-^2$ if $p_2\xi_1 > 0$. A switch occurs when $p_2\xi_1$ changes sign.

Depending on whether $u_1=\omega_+^2$ or $u_1=\omega_-^2$, Hamilton's equations  in the respective $\tau$ intervals become
\ba \label{4.4}
\dot p_1 &= \omega_\pm^2\, p_2, \hspace{2cm}     \dot p_2 = -p_1 \nonumber\\
\dot p_3 &=0, \hspace{2.655cm}      \dot p_4 = -p_3+1.
\end{align}
 Between switches of $u_1$ the solutions are of the form
\ba \label{4.5}
 p_2(\tau) &=A_\pm \cos\omega_\pm\tau + B_\pm\sin\omega_\pm\tau = C_\pm \sin(\omega_\pm \tau - \varphi_\pm) \\
  p_1  &=-\dot p_2,~~~~ p_3 =c_3, ~~~~ p_4 =( -c_3+1)\,\tau + c_4 \nonumber 
\end{align}
where $c_3$, $c_4$, $C_\pm$ are constants, and  $A_\pm $, $B_\pm$,  $\varphi_\pm$ are constants which may dependent on the respective interval. 
If $p_2(\tau) \equiv 0$ in some interval then it is zero everywhere because it cannot be joined continuously to the a nonzero $p_2$ from Eq. (\ref{4.5}).

Since $\omega(\tau) $ is symmetric there must be  intervals of equal length with $\omega(\tau)=\omega_+$  directly to the left and right of $\tau = 0$ (or  $\omega_-$  intervals, but this will not be optimal as shown later). Hence one must have $\varphi_+= 0$ in this interval since then there are  switches in $\omega(\tau)$ at $\tau=\pm \pi/\omega_+$  because $p_2\xi_1 $ vanishes there. It also vanishes at $\tau=0$ but does not change sign because of anti-symmetry of $\xi_1$ and $p_2$ so that $\omega$ has no switch  at $\tau=0$ although $u$ does. Thus $p_2$ is of the form
\ba\label{4.6}
 p_2(\tau)=B_+ \sin(\omega_+\tau)
\end{align}
in the interval $-\pi/\omega_+ \leq \tau \leq \pi/\omega_+$.

\begin{figure}[h]
\begin{center}
%\scalebox{0.26}[0.26]{\includegraphics{time-om.eps}}
  \scalebox{0.9}[0.9]{\includegraphics{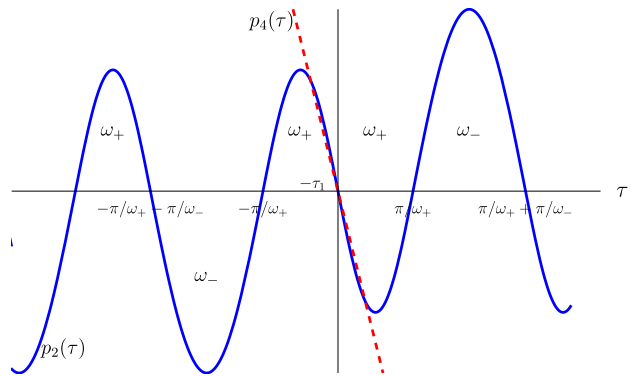}}
\caption{\label{p2om+om-}Solid: $p_2(\tau)$ with symmetric $\omega_\pm$ sequence. Dashed: $p_4(\tau)$. }
\end{center}
\end{figure}

To the left of $\tau=-\pi/\omega_+$ there is an interval with $\omega_-$, then again an $\omega_+$  interval and so on, and similarly to the right of $\tau=\pi/\omega_+$.
Since $p_2(\tau)$ is differentiable different parts of $p_2$ have to be joined accordingly. This yields an anti-symmetric $p_2$  as typically  displayed in Fig. \ref{p2om+om-}.

The procedure for the determination of $\tau_1$ uses the time-development of $\xi_1$ and depends on the interval in which $\frac{1}{2}\tau_{\rm f}$ lies. This will be exemplified for $ \frac{1}{2}\tau_{\rm f} \leq \pi/\omega_+ +\pi/\omega_- $.When $\frac{1}{2}\tau_{\rm f} \leq \pi/\omega_+$ the situation is the same as in Section \ref{section3} and $\tau_1$ is given by Eq. (\ref{tau1}), with $ \omega $ replaced by $\omega_+$.

When $\pi/\omega_+<\frac{1}{2}\tau_{\rm f} \leq \pi/\omega_++\pi/\omega_- $ we calculate  $\xi_1(\tau_{\rm f}/2)$ and $\dot\xi_1(\tau_{\rm f}/2)$ from  $\xi_1(0)$ and $\dot\xi_10)$. By anti-symmetry one has   $\xi_1(0) = 0$ and we put $\dot\xi_1(0)= \lambda$, the exact value  of which will not be needed. Using Eq. (\ref{complex}) one obtains
\ba\label{eta3}
\eta_1 &\equiv \xi_1(\tau_1) +\frac {\rm i}{\omega_+} \dot\xi_1(\tau_1)\nonumber\\
&= \exp[-{\rm i}\omega_+(\tau_1 - 0)](\frac {\rm i}{\omega_+}\lambda+\frac {1}{\omega_+^2})- \frac {1}{\omega_+^2}\nonumber\\
\eta_2 &\equiv \xi_1(\pi/\omega_+) +\frac {\rm i}{\omega_+} \dot\xi_1(\pi/\omega_+)\nonumber\\
=&\exp[-{\rm i}\omega_+(\frac{\pi}{\omega_+} -\tau_1)]\{\Re\eta_1+\frac{\rm i}{\omega+}\omega_+ \Im\eta_1 - \frac{1}{\omega_+^2}\} +
\frac{1}{\omega_+^2}
\nonumber\\
\tilde\eta_3 &\equiv \xi_1(\tau_{\rm f}/2) +\frac {\rm i}{\omega_-} \dot\xi_1(\tau_{\rm f}/2)\nonumber\\
=& \exp[-{\rm i}\omega_-(\tau_{\rm f}/2 -\frac{\pi}{\omega_+})]\{\Re\eta_2 + \frac {\rm i}{\omega_-} \omega_+\Im \eta_2  - \frac {1}{\omega_-^2}\}+ \frac {1}{\omega_-^2}\nonumber\\
\end{align}
By the boundary conditions at $ \frac{1}{2}\tau_{\rm f} $ one has $\tilde\eta_3 = 0$, and thus
\ba\label{eta3a}
0= \Re \eta_2  + \frac {\rm i}{\omega_-} \omega_+\Im \eta_2  - \frac {1}{\omega_-^2} + \exp[{\rm i}\omega_-(\tau_{\rm f}/2  -\frac{\pi}{\omega_+})]\frac {1}{\omega_-^2}.
\end{align}
Taking the real part of this one obtains after a short calculation
\ba\label{tau1a}
\cos[\omega_+\tau_1]=\frac{\omega_+^2}{2\omega_-^2}\{1 +  \cos(\omega_-\tau_{\rm f}/2+\frac{\omega_+ - \omega_-}{\omega_+}\pi)\}.
\end{align}
The l.h.s. cannot exceed 1, while 
the r.h.s. becomes 1 for $\tau_{\rm f}= \tau_{\rm opt}$ where
\ba\label{taufa} 
\tau_{\rm opt}/2 = \frac{\pi}{\omega_+} + \frac{\pi}{\omega_-}-\frac{2}{\omega_-}\arccos[\frac{\omega_-}{\omega_+}],
\end{align}
which lies between $\pi/\omega_+$ and $\pi/\omega_+ + \pi/\omega_-$. Then $\tau_1 = 0$ and the distance becomes the absolute optimum for this particular $\tau_{\rm f} = \tau_{\rm opt}$.

{\em Example} 3. Let $\omega_-= \omega_+/2$. Then  Eq. (\ref{taufa}) yields $\tau_{\rm opt}/2 = \frac{5}{3}\pi/\omega_+$ and the distance $d/d_0$ becomes $\frac{1}{4} \tau_{\rm opt}^2$. If one considered only $\omega_+$ and the corresponding 
$ \tau_{\rm opt}$, one would have $\omega_+\tau_1 = \arccos[3/4] \neq  0$ and the distance would be less.

How to proceed when  the r.h.s. of Eq. (\ref{tau1a}) is larger than 1?  To answer this question  we recall that $p_2$ has also the trivial solution $p_2(\tau) \equiv 0$. Then there are no restrictions on the choice of $\omega(\tau)$. If one decreases $\omega_+$ on the r.h.s of Eq. (\ref{tau1a}) to $\omega_-$ the r.h.s. becomes less or equal to 1. Hence there must be an intermediate $\omega$, denoted by $\tilde\omega_+ $, such that the r.h.s becomes 1. Hence if one uses $[\omega_-, \tilde\omega_+]$ instead of $[\omega_-, \omega_+]$ one gets a solution for $\tau_1$, namely $\tau_1= 0$, so that the sequence $\omega_-$ and $\tilde\omega_+$ gives the largest distance for the given $\tau_{\rm f}$. This means going over to a sub-interval $[\omega_-,\tilde\omega_+]$ of  $[\omega,\omega_+]$ optimizes the distance in this case. There are many sub-intervals with the same property, as seen further below.

In the case $\pi/\omega_+ + \pi/\omega_-<  \tau_{\rm f}/2 \leq \pi/\omega_+ + \pi/\omega_- +  \pi/\omega_+$, i.e. if one starts with $\omega_+$,  switches to $\omega_-$, and to $\omega_+$ before $\tau=0$, i.e. a sequence {\scriptsize $+-+|+-+$} in  Fig. \ref{p2om+om-}, then $\eta_1$ and $\eta_2$ in Eq. (\ref{eta3}) remain unchanged while in $\eta_3$ one replaces $\tau_{\rm f}/2$ by $\pi/\omega_+ + \pi/\omega_-$ and there is an additional $\eta_4$, 
\ba\label{eta4}
\eta_3& = -\Re \eta_2 +2/\omega_-^2 - \frac {\rm i}{\omega_-}\omega_- \Im \eta_2 \nonumber\\
\eta_4 &\equiv \xi_1(\tau_{\rm f}/2) +\frac {\rm i}{\omega_+} \dot\xi_1(\tau_{\rm f}/2)\nonumber\\
&= \exp[-{\rm i}\omega_+(\tau_{\rm f}/2 -\pi/\omega_+ - \pi/\omega_- )]\nonumber\\
&~~~~~~~~~\{\Re \eta_3 + \frac {\rm i}{\omega_+} \omega_-\Im \eta_3  + \frac {1}{\omega_+^2}\} - \frac {1}{\omega_+^2}.
\end{align}
The condition $\eta_4 = 0$ now gives
\ba\label{tau1b}
\cos\omega_+\tau_1 = \frac{\omega_+^2}{\omega_-^2} -1 + \frac{1}{2}\{1 + \cos(\omega_+ \tau_{\rm f}/2 - \frac{\omega_+ - \omega_-}{\omega_-}\pi)\}.
\end{align}
For complete $\omega_\pm$ intervals the exponentials in Eqs. (\ref{eta3}) and (\ref{eta4}) equal -1 and using this the results are easily generalized. In particular, for the  $\omega_\pm$ sequence {\scriptsize$ -+-+|+-+-$}  one obtains
\ba \label{eta5}
\cos(\omega_+\tau_1) =   \frac{\omega_+^2}{\omega_-^2} -1 + \frac{\omega_+^2}{2\omega_-^2}\{1 + \cos(\omega_- \tau_{\rm f}/2 -2\pi\frac{\omega_-}{\omega_+} )\}.
\end{align}

{\em Time optimization}. These results will now be applied to the original problem in which a distance, now denoted by $d_0$, is fixed and the shortest transport time for given $\Omega_\pm$  is sought. If this  $d_0$ is taken  for the definition of the
 scaled variables, $d_0$  becomes $\xi_3(\tau_{\rm f}/2) = 1$.
 The absolutely shortest possible time, $
 \tau_{\rm abs}$, and corresponding $\omega_{\rm res}$ is then, by Example 2, given by
\beq
\tau_{\rm abs} = 2~~~~~~~~~~~~~~~~~~\omega_{\rm res} = 2\pi.
\eeq
 From Fig.~2  the distance traveled in time $\tau_{\rm f}$ is $\frac{1}{4}\tau_{\rm f}^2 -2\tau_1^2$ and if $\tau_{\rm f}$ is to be optimal it must satisfy
\beq\label{d0}
1 = \frac{1}{4}\tau_{\rm f}^2 -2\tau_1^2
\eeq
where $\tau_{\rm f}= \tau_{\rm f}(\omega_-,\omega_+)$.
For given $\omega_\pm$ one obtains $\tau_1$ from Eqs. (\ref{tau1}, \ref{tau1a}, \ref{tau1b}) and generalizations thereof, depending on in which interval  the as yet unknown $\tau_{\rm f}/2$ lies. If $\omega_{\rm res}$ or an integer multiple $n$ thereof lies in [$\omega_-,~\omega_+$] one chooses $\omega(\tau) \equiv n\omega_{\rm res}$ and obtains the absolute optimal $\tau_{\rm abs}$. Different case of increasing complexity will now be discussed.

\begin{figure}[h]
  \begin{center} 
%\scalebox{0.26}[0.26]{\includegraphics{time-om.eps}}
    \scalebox{0.9}[0.9]{\includegraphics{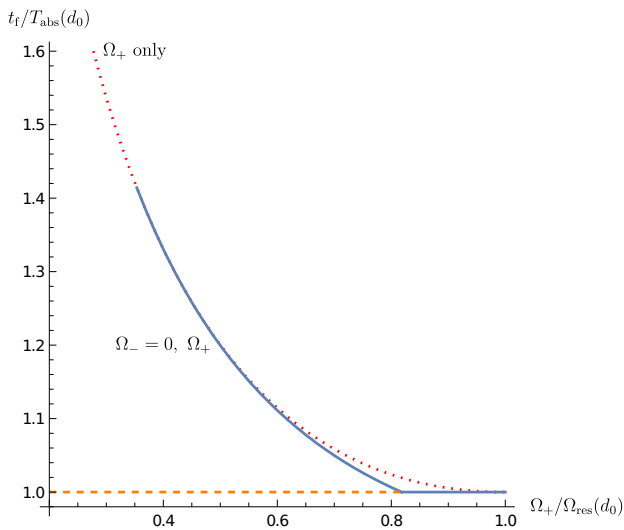}} 
    \caption{\label{ex4}Shortest transport time $t_{\rm f}$ for fixed distance $d_0$, $\Omega_-=0$ and  $0\leq\Omega_+/\Omega_{\rm res}(d_0)\leq 1$.
      Dotted:  $t_{\rm f}$ for fixed $\Omega_+$ without switch in $\Omega$. Solid:  $ \Omega_+/\Omega_{\rm res}(d_0) > \sqrt{2}/4$; initially $\Omega(t) \equiv 0$ and then a switch to $\Omega_+$. For $1/2 + 1/\pi \leq \Omega_+/\Omega_{\rm res}(d_0) \leq 1$ one  has $T_{\rm abs}(d)$. The switch in $\Omega$ can thus lead to a shorter transport time than for $\Omega_+$ alone. } 
\end{center}
\end{figure}

{\em Case:} $\omega_- = 0$,  $0 < \omega_+ < 2\pi$ and the distance 1. If the spring constant is 0 then in the lab frame the mass point $m$ travels free of force and in the the wagon frame under the inertial force. It can happen that it is optimal to start with $\omega_-$. Then  $m$  initially remains at rest in the lab frame until a switch to $\omega_+$ occurs.
If the time development starts with $\omega_+$ there 
can be no switch to $\omega_-$ because the associated time interval $\pi/\omega_-$ is  infinite. Hence in this case the results of Section \ref{section2} and \ref{section3} apply. From 
Fig.~4 it is seen that $\tau_{\rm f}$ decreases with increasing $\omega_+ < 2\pi$. Since $\tau_{\rm f}/2 \leq \pi/\omega_+$ one has,  for optimality,  $\tau_{\rm f}= 2 \pi/\omega_+$ and $ \tau_1 = 0$, by Eqs. (\ref{2.3},\ref{2.4}). From Eq. (\ref{d0}) one then obtains $\tau_{\rm f}^2 = 4$ so that in this case one must have  $\omega_+ = \pi/\sqrt{2}\equiv \tilde\omega_+$. Thus if $\omega_+ >\tilde\omega_+$  one starts with $\omega_- = 0$ and then there is a switch to $\omega_+$ at some later time. In this case Eq. (\ref{tau1a}) holds for $\tau_1$ and it becomes 0 for $\tau_{\rm f} = \tau_{\rm opt}  $ given by Eq. (\ref{taufa}). Taking the limit $\omega_- \to 0$  one finds $\tau_{\rm opt}= (2\pi  + 4)/\omega_+ $. This must equal $\tau_{\rm abs} = 2$ which gives $\omega_+ = \pi +2 \equiv \omega_{\rm abs}$. From this value of $\omega_+$ on one obtains the absolute time minimum.  The optimal time as a function of $\omega_+$ is displayed in Fig. \ref{ex4}.

{\em Protocol.} This  depends on $\omega_+$ and is as in Section \ref{section2} when $\omega_+\leq \tilde\omega_+$. When $ \tilde\omega_+ < \omega_+ \leq \omega_{\rm abs}  $ one determines $\tau_{\rm f}$ and $\tau_1$ from Eqs. (\ref{tau1b}) and (\ref{d0}), starts  with $\omega_- =0$ for the time duration $ -\pi/\tilde\omega_+ +\tau_{\rm f}/2$ and with $u=1$, then switches to $\omega_+$ and continues for the time $-\tau_1 + \pi/\tilde\omega_+$, then switches to $u=-1$ for the time $\tau_1$ and continues by symmetry, resp. anti-symmetry. When $\omega_{\rm abs}=2+\pi< \omega_+ \leq \omega_{\rm res}$ one chooses the protocol for $\omega_+ = \omega_{\rm abs}$.

\begin{figure}[h]  
  \begin{center}  
%\scalebox{0.26}[0.26]{\includegraphics{time-om.eps}}
    \scalebox{0.9}[0.9]{\includegraphics{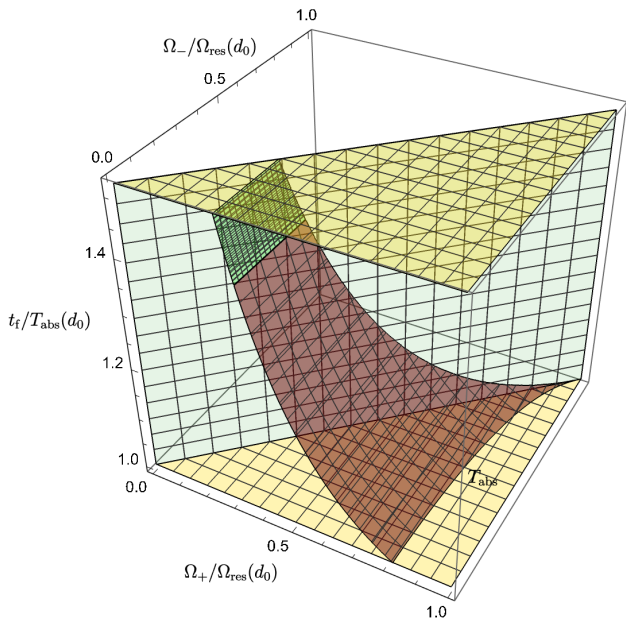}} 
    \caption{\label{om+om-}Shortest transport time $t_{\rm f}$ for fixed distance $d_0$ and  $0\leq\Omega_-/\Omega_{\rm res}(d_0)  \leq\Omega_+/\Omega_{\rm res}(d_0)\leq 1$. For $\Omega_+  /\Omega_{\rm res}(d_0)\leq  \pi/\sqrt{2} $ there is only $\Omega_+$  and no switch (close meshed region). For $\{\Omega_-,\Omega_+\}$ in the region denoted by $T_{\rm abs}$ at the r.h.s. one has the shortest time $T_{\rm abs}$. The intersection of the surface with the front plane is the curve of Fig.~ 9 and that with the diagonal plane is the left part of the curve of Fig.~4 until 1.}
    \end{center}
  \end{figure}

{\em Case:} $0< \omega_- < \omega_+ < \omega_{\rm res} = 2\pi$. As in the preceding case,  only $\omega_+$ is relevant if $\omega_+ \leq \tilde\omega_+ = \pi/\sqrt{2} $. Then $\tau_{\rm f}(\omega_-,\omega_+)/2 \leq \pi/\omega_+$ and is independent of $\omega_-$.  This is the upper close meshed region in Fig.~10. For  $\omega_+> \tilde\omega_+$ there are on the l.h.s. of Fig.~8  two or more alternating $\omega_\pm$'s  for the time development. If there  are two, one starts with $\omega_-$, and the initial time $-\tau_{\rm f}/2$  satisfies $\pi/\omega_+ \leq \tau_{\rm f}/2 \leq \pi/\omega_+ + \pi/\omega_- $. In this case  Eqs.  (\ref{d0}) and (\ref{tau1a})   apply. If the  l.h.s. of Eq. (\ref{tau1a}) is less or equal to 1  then  one can determine $\tau_1$ and $\tau_{\rm f}(\omega_-,\omega_+)$, displayed by the coarse meshed region in Fig.  \ref{om+om-}. Putting   $\cos[\omega_+\tau_1]= 1$ one obtains with $\tau_{\rm f}= \tau_{\rm abs}= 2\pi$ from Eq. (\ref{tau1a}) the boundary curve at the bottom of  the coarse meshed surface which borders the region denoted by $T_{\rm abs}$. In this region  there is no solution for $\tau_1$. As before, here the solution $p_2(\tau)\equiv 0$ can be used and then there are no restrictions on $\omega(\tau)$.  If one starts from the point $\{\omega_-,\omega_+\}$ and first decreases $\omega_+$ until one hits the boundary curve and then similarly increases $\omega_-$ one obtains the end points of an arc on the boundary curve. Every point $\{\hat\omega_-,\hat\omega_+\}$ on this arc  satisfies $\{\omega_- \leq \hat\omega_-\leq \hat\omega_+ \leq \omega_+ \}$ and yields 
$\tau_{\rm abs}$. Thus there is again an improvement over the single $\omega_+$ case. 

If there were a third, preceding, interval, i.e. with $\omega_+$, then $\tau_{\rm f}(\omega_-,\omega_+)/2 >\pi/\omega_+ + \pi/\omega_- $ and $\tau_{\rm f}$ would thus be larger than that with only two periods. Hence a third period does not occur. By a similar calculation, interchanging $\omega_+$ and  $\omega_-$  leads to a larger transport time.

{\em Protocol}:  When $\omega_+\leq \tilde\omega_+ =
\pi/\sqrt{2}$ one proceeds with $\omega_+$ as in Section \ref{section2}.  When $ \omega_+>\tilde\omega_+ $ one determines $\tau_{\rm f}(\omega_-,\omega_+)$ and $\tau_1$ from Eqs. (\ref{tau1a}) and (\ref{d0}), provided a solution for $\tau_1$ exists. Then one has an $\omega_\pm$ sequence of the form {\scriptsize $ -+|+-$} and thus one starts  with $u=1$ and  $\omega_- $ from time $-\tau_{\rm f}/2$ to time $ -\pi/\omega_ +$ where one  switches to $\omega_+$. Then one continues until time $-\tau_1$, where one switches to $u=-1$   and continues to $\tau=0$ where there is a switch back to $u=1$. For $\tau>0$ one continues by symmetry, resp. anti-symmetry. When there is no solution for $\tau_1$, i.e when the point $\{\omega_-,\omega_+\}$ lies in the region denoted by $T_{\rm abs}$ in Fig.~10, then one can choose a protocol for any point on the above  arc. This will yield $\tau_{\rm abs}$  and in this case the protocol is not unique.

{\em Case:} $\omega_{\rm res} = 2\pi \leq \omega_- < \omega_+ <2 \,\omega_{\rm res}$. Arguing as before, one has {\scriptsize $+-+|+-+$} and  {\scriptsize $-+-+|+-+-$} as possible $\omega_\pm$ sequences. To the first sequence Eq. (\ref{tau1b}) applies and to the second Eq. (\ref{eta5}). One now solves Eq. (\ref{d0}) together with Eq. (\ref{tau1b}) for $\tau_{\rm f}$ under the condition that$\tau_{\rm f}/2$ lies in the last $\omega_+$ interval. In Fig.~11 this gives the left surface outside of which there is no solution for $\tau_1$. In a similar way one obtains the right surface for the second sequence.  On the boundary curve at the bottom  one has  $\tau_{\rm abs}$ and the curve is obtained from $\cos(\omega_+\tau_1) =1$. The two $\omega_\pm$ sequences are separated by the dashed curve under the surface. This curve is obtained by putting $\tau_{\rm f}/2 =2\pi/\omega_+ + \pi/\omega_-$ in Eqs. (\ref{tau1b},~\ref{d0}).
Its end point on the boundary curve  is given by $\{\frac{1}{2}+\frac{1}{2}\sqrt{2},1+\frac{1}{2}\sqrt{2}\}\,\omega_{\rm res}$ and on the diagonal by $\frac{1}{4}\sqrt{34}\,\omega_{\rm res}$.

In the region denoted by $T_{\rm abs}$ there is no solution for $\tau_1$. Again one can choose any point $\{\hat\omega_-,\hat\omega_+\}$ on the arc constructed as before to obtain $\tau_{\rm abs}$. Reversing the sequence to  {\scriptsize $-+-|-+-$} leads to larger transport times.

{\em Protocol}: If for a given $\{\omega_-,\omega_+\}$ one has $\omega_-\leq (\frac{1}{2}+\frac{1}{2}\sqrt{2})\, \omega_{\rm res}$ or if  a solution 
for $\tau_1$ in Eq. (\ref{tau1b}) exists,  one has a  sequence {\scriptsize $+-+|+-+$}, from Fig.~11. If a solution exists
  the protocol is analogous to the previous case above. If not, one picks a point $\{\hat\omega_-,\hat\omega_+\}$ on the arc on the boundary curve, as before, and uses the protocol for this point with $\tau_{\rm f}= \tau_{\rm abs} $.  Otherwise,  one has a sequence {\scriptsize $\,\,-+-+|+-+-\,\,$} and the procedure is analogous.

\begin{figure}[h] 
  \begin{center} 
%\scalebox{0.26}[0.26]{\includegraphics{time-om.eps}}
    \scalebox{0.9}[0.9]{\includegraphics{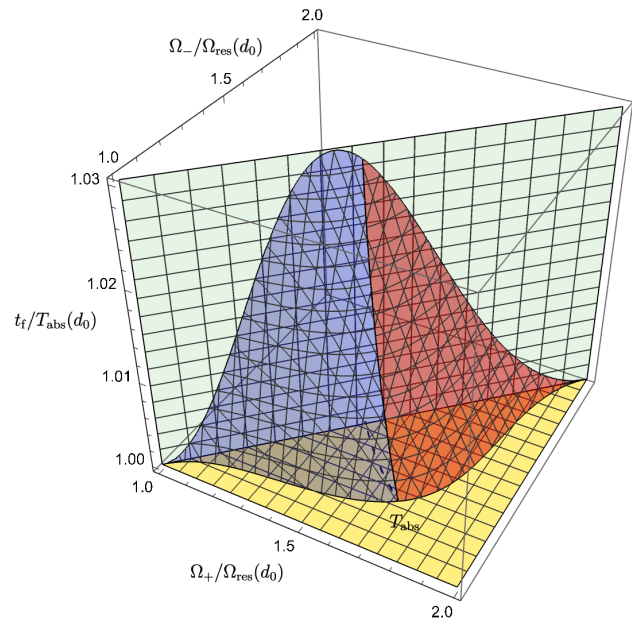}} 
    \caption{\label{om+om-om+}Shortest transport time $t_{\rm f}$ for fixed distance $d_0$ and $1\leq\Omega_-/\Omega_{\rm res}(d_0)  \leq\Omega_+/\Omega_{\rm res}(d0)\leq 2$.
      The left side of the surface belongs to an $\Omega_\pm$ sequence {\scriptsize $+-+|+-+$}, the right side to {\scriptsize $-+-+|+-+-$}, separated by the dashed line in the bottom plane. For $\{\Omega_-,\Omega_+\}$   in the region denoted by $T_{\rm abs}$ one obtains the shortest time $T_{\rm abs}$ by going over to a point on the boundary corresponding to a sub-interval of $[\Omega_-,\Omega_+]$.  }
\end{center}
\end{figure}
  
\section{Summary and Discussion} \label{discussion}

Protocols for the fastest possible transport of a classical harmonic oscillator (h.o.) over a distance $d$ have been derived where both initially and finally everything is at rest, i.e. the position of the h.o. is at rest  and the h.o. is  in its equilibrium position and also at rest. The acceleration $a(t)$ is assumed to satisfy $-a_{\rm max} \leq a(t) \leq a_{\rm max}$. 

First, with fixed h.o. frequency $\Omega$, for the shortest transport time the optimal acceleration alternates between $\pm a_{\rm max} $. It was shown that one starts with $a_{\rm max}$ and that there are   three switches or, for special values  $\Omega=n\Omega_{\rm res}(d) = 2\pi n\sqrt{a_{\rm max}/d}$, $n=1,2,\cdots$, only one switch. The switch times were determined.

The dependence of the shortest transport time, denoted by $t_{\rm f}$, 
on $d$, $\Omega$ and $a_{\rm max}$ was found, cf. Figs.~3  and 4. The optimal time $t_{\rm f}$ is proportional to  $1/\sqrt{a_{\rm max}}$, diverges for $\Omega \to 0$ and, not surprisingly,  for  $\Omega \to \infty$ converges to $T_{\rm abs}(d)= 2\sqrt{d/a_{\rm max}}$, the optimal time for a wagon without h.o.. The function  $t_{\rm f}(d)$ approaches  $T_{\rm abs}(d)$ for large $d$.
 Surprisingly, sometimes  it is advantageous to go {\em backwards} for a while, but not as far back as  the initial position. 

 Second, in addition to $a(t)$ a time-dependent $\Omega(t)$ satisfying $\Omega_- \leq \Omega(t) \leq \Omega_+$ was considered. In this case the behavior of $t_{\rm f}$ depends sensitively on $\Omega_\pm$. If $n\,\Omega_{\rm res}(d)$ lies in the interval $[\Omega_-,\Omega_+]$ for some  $n$ then choosing $n\,\Omega_{\rm res}(d)$ will give the minimal time $T_{\rm abs}(d)$.

 If   $\Omega_+ \leq \frac{1}{2\sqrt{2}} \Omega_{\rm res}$ then $\Omega(t) \equiv \Omega_+$, there is no switch in $\Omega$, and $\Omega_-$ does not enter. Otherwise there are two alternatives if $\Omega_+ < \Omega_{\rm res}$:\\
 ({\em i}) One starts with $\Omega_-$, switches to $\Omega_+ $ and then back to $\Omega_-$.\\
 ({\em ii}) Or  there are $\tilde\Omega_\pm$, depending on  $\Omega_\pm$, with  $\Omega_- \leq \tilde\Omega_-\leq \tilde\Omega_+ \leq \Omega_+$ and one starts with $\tilde\Omega_-$, switches to $\tilde\Omega_+ $ and then back to $\tilde\Omega_-$. In this case one obtains the minimal time $T_{\rm abs}(d)$. In the $\Omega_- - \Omega_+$ plane this happens for $\{\Omega_-,\Omega_+\}$ in a region, cf. Fig.~10.
 
If   $n\,\Omega_{\rm res} < \Omega_-\leq \Omega_+ < (n+1) \Omega_{\rm res}$ the situation is similarly involved and depicted for $n=1$ in Fig.~11 .

The Pontryagin Maximum Principle was employed,  first  for constant $\Omega$ with $a(t)$ as a control variable, and then with $a(t)$  and $\Omega(t)$ as control variables. Symmetry properties played an important role which were proved  for constant $\Omega$ and assumed in an analogous form for time-dependent $\Omega$.

 One may also want to impose restrictions on the velocities $\dot x_{\rm w}$ and  $\dot x_{\rm h}$ or on the relative displacement $x_{\rm h}$ of the h.o.. Within the PMP this may be formulated by means of Lagrangian multipliers. In \cite{stefmuga2011} the relative displacement was assumed to be bounded and taken as the only control. However, in this case  there are $\delta (t)$-like forces  at the time of a switch  acting on the h.o., and no oscillations occur.

The above results for constant $\Omega$ have immediate applications to cranes
for small-angle oscillations of the payload where the the rope length $l$  is constant. For time dependent $l(t)$ modifications are needed since $l(t)$ is not  related to the frequency $\Omega(t)$  in the same way as the spring constant.

The harmonic oscillator considered here is an idealized system. However, it may serve as a benchmark for more realistic models, e.g. if the switches are short but smooth rather than instantaneous.

\end{document}